\documentclass{emulateapj}
\usepackage{apjfonts}

\slugcomment{Submitted 2003 November 17; accepted 2004 February 3}

\shorttitle{STAR FORMATION RATE AND SUPERNOVA NEUTRINOS}
\shortauthors{ANDO}

\begin{document}

\title{Cosmic Star Formation History and the Future Observation \\of
Supernova Relic Neutrinos}

\author{Shin'ichiro Ando}
\affil{Department of Physics, School of Science, University of Tokyo,
7-3-1 Hongo, Bunkyo-ku, Tokyo 113-0033, Japan}
\email{ando@utap.phys.s.u-tokyo.ac.jp}

\begin{abstract}

We investigate the flux and event rate of supernova relic neutrinos
 (SRNs) and discuss their implications for the cosmic star formation
 rate.
Since SRNs are diffuse neutrino background emitted from past
 core-collapse supernova explosions, they contain fruitful information
 on the supernova rate in the past and present universe, as well as on
 the supernova neutrino spectrum itself.
As reference models, we adopt the supernova rate model based on recent
 observations and the supernova neutrino spectrum numerically calculated
 by several groups.
In the detection energy range $E_e>10$ MeV, which will possibly be a
 background-free region in the near future, the SRN event rate is found
 to be 1--2 yr${}^{-1}$ at a water Cerenkov detector with a fiducial
 volume of 22.5 kton, depending on the adopted neutrino spectrum.
We also simulate the expected signal with one set of the reference
 models by using the Monte Carlo method and then analyze these
 pseudodata with several free parameters, obtaining the distribution of
 the best-fit values for them.
In particular, we use a parameterization such that $R_{\rm SN}(z)=R_{\rm
 SN}^0 (1+z)^\alpha$, where $R_{\rm SN}(z)$ is the comoving supernova
 rate density at redshift $z$ and $R_{\rm SN}^0$ and $\alpha$ are free
 parameters, assuming that the supernova neutrino spectrum and
 luminosity are well understood by way of a future Galactic supernova
 neutrino burst or the future development of numerical supernova
 simulations.
The obtained $1\sigma$ errors for these two parameters are found to be
 $\delta\alpha /\langle\alpha\rangle =30\%~(7.8\%)$ and $\delta R_{\rm
 SN}^0/\langle R_{\rm SN}^0\rangle =28\%~(7.7\%)$ for a detector with
 an effective volume of 22.5 kton\,$\times$\,5 yr (440 kton\,$\times$\,5
 yr), where one of the parameters is fixed.
On the other hand, if we fix neither of the values for these two
 parameters, the expected errors become rather large, $\delta\alpha
 /\langle\alpha\rangle =37\%$ and $\delta R_{\rm SN}^0/\langle R_{\rm
 SN}^0 \rangle =55\%$, even with an effective volume of
 440 kton\,$\times$\,5 yr.

\end{abstract}


\section{INTRODUCTION}
\label{sec:Introduction}

In recent years we have made remarkable progress in our knowledge
concerning how the cosmic star formation history proceeded in the past
and concerning the fraction of baryons locked up in stars and gas in
the local universe.
These points were inferred from observations of the light emitted by
stars of various masses at various wavelengths.
\citet{Madau96} investigated the galaxy luminosity density of rest-frame
ultraviolet (UV) radiation up to $z\sim 4$, and they converted it into
the cosmic star formation rate (SFR).
The rest-frame UV light is considered to be a direct tracer of star
formation because it is mainly radiated by short-lived massive stars.
After the pioneering study by \citeauthor{Madau96}, a wealth of data
have become available in the form of the cosmic SFR in a wide range of
redshifts; these data were inferred from observations using far-infrared
(FIR)/submillimeter dust emission \citep{Hughes98,Flores99} and
near-infrared (NIR) H$\alpha$ line emission
\citep{Gallego95,Gronwall98,Tresse98,Glazebrook99}, as well as the
rest-frame UV emission from massive stars (\citealt{Lilly96};
\citealt{Cowie96}; \citealt{Connolly97}; \citealt*{Sawicki97};
\citealt{Treyer98}; \citealt*{Madau98a}; \citealt*{Pascarelle98};
\citealt{Steidel99}).

In these traditional methods, however, there are a fair number of
ambiguities when the actual observables are converted into the cosmic
SFR \citep*{Somerville01}.
First, observable samples are generally flux-limited, and thus the
intrinsic luminosity of the faintest objects in the sample changes with
redshift.
In order to understand the true redshift dependence of the total
luminosity density, this incompleteness is generally corrected by using
a functional form (i.e., a Schechter function) of the luminosity
function obtained from the observations themselves.
Unfortunately, since the luminosity function is not well established
observationally (especially for high-$z$ regions), it is uncertain
whether the Schechter-function fit is good enough or not.
Second, the conversion from luminosity density to SFR generally relies
on stellar population models, an assumed star formation history, and
an initial mass function (IMF), which are not also well established
yet.
Finally, if the tracer of star formation is an optical or UV luminosity,
then the effects of dust extinction are nonnegligible.
Although this problem is less critical in other wave bands such as
NIR H$\alpha$ or FIR/submillimeter, the bulk of current data consists of
rest-frame UV observations, especially of high-redshift regions.
After adopting some correction law for dust extinction, the rest-frame UV
data become rather consistent with H$\alpha$ or submillimeter data
points; still, in this case it is unknown whether the UV and
submillimeter sources are identical, which is very important for
measuring the cosmic SFR.

Thus, our knowledge concerning the cosmic SFR is quite crude, and
therefore another type of observation that is independent of the above
methods would be very important.
In this paper we consider supernova relic neutrinos (SRNs), i.e., a
diffuse background of neutrinos that were emitted from past supernova
explosions.
Type Ib, Ic, and II supernova explosions are considered to have traced
the cosmic SFR, because they are directly connected with the death of
massive stars with $M\gtrsim 8M_\sun$ and their lifetime is expected to
be very short compared with the time scale of star formation.
These events are triggered by gravitational collapse, and 99\% of the
gravitational binding energy is released as neutrinos; this basic
scenario was roughly confirmed by the well-known observation of the
neutrino burst from SN 1987A by the Kamiokande II and IMB detectors
\citep{Hirata87,Bionta87}.
The advantages of using SRNs to probe the cosmic SFR are as follows:
First, as mentioned above, if we can probe the supernova rate from SRN
observation, it can be directly transformed to the SFR assuming the
IMF, because supernovae are short-lived astrophysical events.
The second advantage, which is more important, is that neutrinos
are completely free of dust extinction.
This point is the same with observations in the submillimeter wave band;
however, neutrinos are emitted directly from stars, whereas
submillimeter radiation comes from dust and is an indirect process.

The SRN flux and the event rates at a currently working large-volume
water Cerenkov detector, Super-Kamiokande (SK), have been investigated
by many researchers using theoretically/observationally modeled SFRs
\citep*{Totani95,Totani96a,Malaney97,Hartmann97,Kaplinghat00,Ando03a}.
More recently, the SK collaboration obtained a 90\% CL upper limit
on the SRN flux, i.e., 1.2 cm${}^{-2}$ s${}^{-1}$ in the energy range
$E_\nu >19.3$ MeV \citep{Malek03}.
This severe constraint is only about factor of 3--6 larger than the
typical theoretical models and is very useful for obtaining several
rough estimations of the cosmic SFR \citep{Fukugita03,Strigari03} and
probing the properties of neutrinos as elementary particles
\citep{Ando03c,Ando03f}.

However, we need a further $\sim 40$ years to reduce the current limit
by a factor of 3 if we use the SK detector with current performance.
This is because there is no energy window for SRN detection where the
SRN signal dominates other background events coming from various
sources, such as solar, reactor, and atmospheric neutrinos, as well as
cosmic-ray muons \citep{Ando03a}.
Therefore, current observations are seriously affected by the other
backgrounds and take much time to reach the required sensitivity.
In order to overcome this difficulty, a very interesting and promising
method was proposed to directly tag electron antineutrinos
($\bar\nu_e$), and it is now in the research and development phase
\citep{Beacom03b}.
The basic idea is to dissolve 0.2\% gadolinium trichloride (GdCl$_3$)
into the pure water of SK.
With this mixture, 90\% of the neutrons produced by the $\bar\nu_ep\to
e^+n$ reaction are captured on Gd and then decay with 8 MeV gamma
cascades.
When we detect these gamma cascades, as well as the preceding Cerenkov
radiation from positrons, it indicates that these signals come from
original $\bar\nu_e$, not from other flavor neutrinos or muons.
With this method, we can remove the background signals in the energy
range 10--30 MeV, in which before removal there is a huge amount of
background from solar neutrinos ($\nu_e$) and atmospheric muon-neutrinos
($\nu_\mu,\bar\nu_\mu$) or cosmic-ray muon induced events.
Because the expected SRN rate is estimated to be 1--2 yr${}^{-1}$ in the
energy range 10--30 MeV, the Gd-loaded SK detector (Gd-SK) would enable
us to detect a few SRN events each year.

Therefore, it is obviously important and urgent to make a detailed
investigation of the performance of such future detectors.
In this paper we focus on how far we can probe the cosmic supernova
rate by SRN observations at Gd-SK and at the hypothetical Gd-loaded
Hyper-Kamiokande (Gd-HK) detector or Gd-loaded Underground Nucleon Decay
and Neutrino Observatory (Gd-UNO).
Because the expected event rate of SRNs is about 1--2 yr${}^{-1}$ in the
detectable energy range (10--30 MeV) using a detector with the size of
SK, it would be quite difficult to obtain the spectral information of
SRNs, even if we observed for 5 years.
On the other hand, with the currently proposed megaton-class detectors
such as HK and UNO, we can expect to obtain a great deal of information
about the SRN spectrum, which will be useful for inferring the SFR-$z$
relation.
Using the Monte Carlo (MC) method, we simulate an expected SRN signal at
these future detectors, and then we analyze these hypothetical data with
a few free parameters and discuss implications from future SRN
observations.

This paper is organized as follows.
In \S~\ref{sec:Formulation and Models} we give the formulation for
calculating the SRN flux and discuss several models that are adopted in
our calculations, and in \S~\ref{sec:Flux and Event Rate of Supernova
Relic Neutrinos} we show the results of our calculation with some
reference models.
In \S~\ref{sec:Monte Carlo Simulation for Future Detector Performance}
the MC simulation of the expected signal at the future Gd-loaded
detectors, which is generated from the reference models, is presented,
and then we analyze these hypothetical data using several free
parameters concerning the cosmic SFR.
Finally, we discuss other possibilities in \S~\ref{sec:Discussion}.

\section{FORMULATION AND MODELS}
\label{sec:Formulation and Models}

\subsection{Formulation}
\label{sub:Formulation}

The present number density of SRN ($\bar\nu_e$), whose energy is in the
interval $E_\nu\sim E_\nu +dE_\nu$, emitted in the redshift interval
$z\sim z+dz$, is given by
\begin{eqnarray}
 dn_\nu(E_\nu)&=&R_{\rm SN}(z)(1+z)^3\frac{dt}{dz}
  dz\frac{dN_\nu (E_\nu^\prime)}{dE_\nu^\prime}dE_\nu^\prime
  (1+z)^{-3}\nonumber\\
 &=&R_{\rm SN}(z)\frac{dt}{dz}dz\frac{dN_\nu (E_\nu^\prime)}
   {dE_\nu^\prime}(1+z)dE_\nu ,
   \label{eq:dn_nu}
\end{eqnarray}
where $E_\nu^\prime =(1+z)E_\nu$ is the energy of neutrinos at redshift
$z$, which is now observed as $E_\nu$; $R_{\rm SN}(z)$ represents
the supernova rate per comoving volume at $z$, and hence the factor
$(1+z)^3$ should be multiplied to obtain the rate per physical volume at
that time; $dN_\nu /dE_\nu$ is the number spectrum of neutrinos emitted
by one supernova explosion; and the factor $(1+z)^{-3}$ comes from the
expansion of the universe.
The Friedmann equation gives the relation between $t$ and $z$ as
\begin{equation}
\frac{dz}{dt}=-H_0(1+z)\sqrt{(1+\Omega_mz)(1+z)^2-\Omega_\Lambda
 (2z+z^2)},
\label{eq:dz_dt}
\end{equation}
and we adopt the standard $\Lambda$CDM cosmology ($\Omega_m=0.3,
\Omega_\Lambda =0.7,$ and $H_0=70~h_{70}~\mathrm{km~s^{-1}~
Mpc^{-1}}$).\footnote{Although we use the specific cosmological model
here, the SRN flux itself is completely independent of such cosmological
parameters, as long as we use observationally inferred SFR models (see
their cancellation between eqs. [\ref{eq:SRN flux}] and [\ref{eq:SFR}]),
as already discussed in \citet{Ando03a}.}
We now obtain the differential number flux of SRNs, $dF_\nu /dE_\nu$,
using the relation $dF_\nu /dE_\nu =c\,dn_\nu /dE_\nu$:
\begin{eqnarray}
 \frac{dF_\nu}{dE_\nu}&=&\frac{c}{H_0}\int_0^{z_{\rm max}}
  R_{\rm SN}(z)\frac{dN_\nu (E_\nu^\prime)}{dE_\nu^\prime}
  \nonumber\\
 &&{}\times\frac{dz}{\sqrt{(1+\Omega_mz)(1+z)^2-\Omega_\Lambda
  (2z+z^2)}},
  \label{eq:SRN flux}
\end{eqnarray}
where we assume that gravitational collapses began at the redshift
$z_{\rm max}=5$.

\subsection{Model for Cosmic Star Formation Rate}
\label{sub:Models for Cosmic Star Formation Rate}

\begin{figure}[htbp]
\begin{center}
\includegraphics[width=8.5cm]{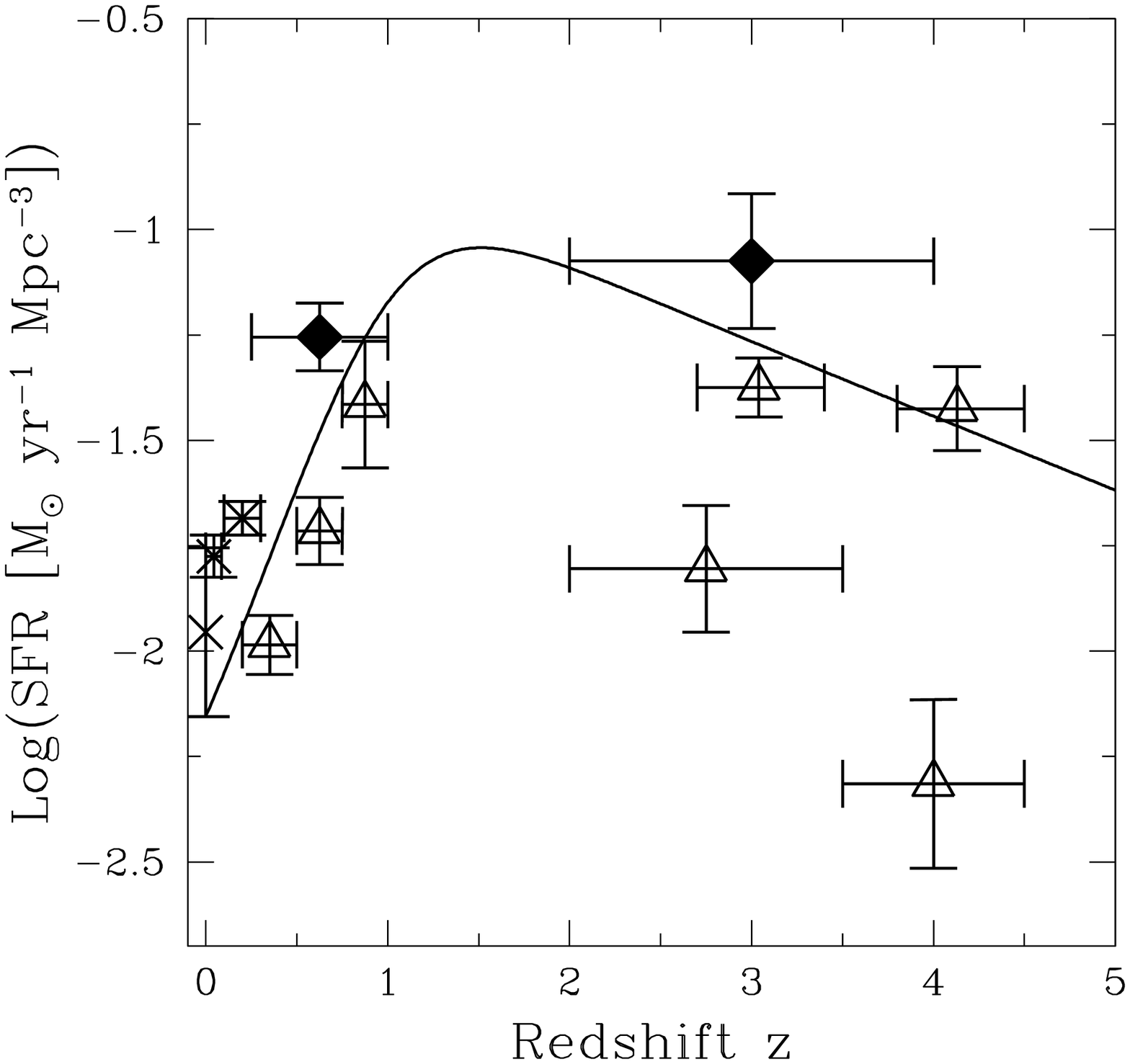}
\caption{Cosmic star formation rate as a function of redshift. Data
 points are given by rest-frame UV ({\it open triangles}; \citealt{Lilly96};
 \citealt{Madau96}; \citealt{Steidel99}), NIR H$\alpha$ ({\it crosses};
 \citealt{Gallego95}; \citealt{Gronwall98}; \citealt{Tresse98}), and
 FIR/submm ({\it filled diamonds}; \citealt{Flores99}; \citealt{Hughes98})
 observations. The solid curve represents our reference model given by
 eq. (\ref{eq:SFR}). The standard $\Lambda$CDM cosmology is adopted
 ($\Omega_m=0.3,\Omega_\Lambda =0.7, H_0=70$ km s${}^{-1}$ Mpc${}^{-1}$)
 \label{fig:SFR}}
\end{center}
\end{figure}

As our reference model for the SFR, we adopt a model that is based on
recent progressive results of rest-frame UV, NIR H$\alpha$, and
FIR/submillimeter observations; a simple functional form for the SFR per
unit comoving volume is given as \citep{Porciani01}
\begin{eqnarray}
 \psi_\ast (z) &=&0.32h_{70}
  \frac{\exp (3.4z)}{\exp (3.8z)+45}~M_\sun~\mathrm{yr^{-1}~Mpc^{-3}}
  \nonumber\\
 &&{}\times\frac{\sqrt{(1+\Omega_mz)(1+z)^2-\Omega_\Lambda (2z+z^2)}}
  {(1+z)^{3/2}}.
  \label{eq:SFR}
\end{eqnarray}
Figure \ref{fig:SFR} shows the SFR $\psi_\ast (z)$ with the various data
points from rest-frame UV \citep{Lilly96,Madau96,Steidel99}, H$\alpha$
line \citep{Gallego95,Gronwall98,Tresse98}, and FIR/submillimeter
\citep{Flores99,Hughes98} observations; these data points are not
corrected for dust extinction.
In the local universe, all studies show that the comoving SFR
monotonically increases with $z$ out to a redshift of at least 1.
Our reference model (eq. [\ref{eq:SFR}]) is consistent with mildly
dust-corrected UV data at low redshift; on the other hand, it may
underestimate the results of the other wave band observations.
In our previous paper \citep{Ando03a}, we investigated the dependence on
the several adopted SFR models, which were only different at
high-redshift regions ($z\gtrsim 1.5$); our reference model
(\ref{eq:SFR}) was referred to as the ``SF1'' model there.
We showed that the SRN flux at $E_\nu >10$ MeV is highly insensitive to
the difference among the SFR models (owing to the energy redshift, as
discussed in \S~\ref{sub:Flux of Supernova Relic Neutrinos}), and
therefore we do not repeat such discussions in the present paper.

We obtain the supernova rate ($R_{\rm SN}(z)$) from the SFR by assuming
the Salpeter IMF ($\phi (m)\propto m^{-2.35}$) with a lower cutoff
around $0.5M_\sun$, and that all stars with $M>8M_\sun$ explode as
core-collapse supernovae, i.e.,
\begin{eqnarray}
 R_{\rm SN}(z)&=&\frac{\int_{8M_\sun}^{125M_\sun}dm~\phi(m)}
  {\int_{0}^{125M_\sun}dm~m\phi(m)}\psi_\ast (z)\nonumber\\
 &=&0.0122M_\sun^{-1}\psi_\ast (z).
  \label{eq:SN rate}
\end{eqnarray}
The resulting local supernova rate agrees within errors with the
observed value of $R_{\rm SN}(0)=(1.2\pm 0.4)\times 10^{-4}h_{70}^3
\mathrm{~yr^{-1}~Mpc^{-3}}$ (e.g., \citealt*{Madau98b} and references
therein).
In fact, the totally time-integrated neutrino spectrum from massive
stars ($\gtrsim 30M_\sun$) could be very different from the models that
we use (and give in the next subsection), possibly because of, e.g.,
black hole formation.
However, the conversion factor appearing in equation (\ref{eq:SN rate})
is highly insensitive to the upper limit of the integral in the
numerator; for instance, if we change the upper limit in the numerator
to $25M_\sun$, the factor becomes $0.010M_\sun^{-1}$, which is only
slightly different from the value in equation (\ref{eq:SN rate}).

\subsection{Neutrino Spectrum from Supernova Explosions}
\label{sub:Neutrino Spectrum from Supernova Explosions}

For the neutrino spectrum from each supernova, we adopt three reference
models by different groups, i.e., simulations by the Lawrence Livermore
(LL) group \citep{Totani98} and Thompson, Burrows, \& Pinto (2003,
hereafter TBP), and the MC study of spectral formation by Keil, Raffelt,
\& Janka (2003, hereafter KRJ).
In this field, however, the most serious problem is that the recent
sophisticated hydrodynamic simulations have not obtained the supernova
explosion itself; the shock wave cannot penetrate the entire core.
Therefore, many points still remain controversial, e.g., the average
energy ratio among neutrinos of different flavors, or how the
gravitational binding energy is distributed to each flavor.
All these problems are quite serious for our estimation, since the
binding energy released as $\bar\nu_e$ changes the normalization of the
SRN flux, and the average energy affects the SRN spectral shape.
Thus, we believe that these three models from different groups will be
complementary.

The numerical simulation by the LL group \citep{Totani98} is considered
to be the most appropriate for our estimation, because it is the only
model that succeeded in obtaining a robust explosion and in calculating
the neutrino spectrum during the entire burst ($\sim 15$ s).
According to their calculation, the average energy difference between
$\bar\nu_e$ and $\nu_x$, where $\nu_x$ represent the nonelectron-flavor
neutrinos and antineutrinos, was rather large and the complete
equipartition of the binding energy was realized
$L_{\nu_e}=L_{\bar\nu_e}=L_{\nu_x}$, where $L_{\nu_\alpha}$ represents
the released gravitational energy as $\alpha$-flavor neutrinos.
The neutrino spectrum obtained by their simulation is well fitted by a
simple formula, which was originally given by KRJ as
\begin{equation}
 \frac{dN_\nu}{dE_\nu}=\frac{(1+\beta_\nu)^{1+\beta_\nu}L_\nu}
  {\Gamma (1+\beta_\nu)\bar E_\nu^2}
  \left(\frac{E_\nu}{\bar E_\nu}\right)^{\beta_\nu}
  e^{-(1+\beta_\nu)E_\nu/\bar E_\nu},
  \label{eq:beta fit}
\end{equation}
where $\bar E_\nu$ is the average energy; the values of the fitting
parameters for the $\bar\nu_e$ and $\nu_x$ spectrum are summarized in
Table \ref{table:fitting parameters}.
\begin{deluxetable*}{crrrrrrrc}[htbp]
\tabletypesize{\scriptsize}
\tablecaption{Fitting Parameters for Supernova Neutrino
 Spectrum \label{table:fitting parameters}}
\tablehead{
 & \colhead{Mass} & \colhead{$\bar E_{\bar\nu_e}$}
 & \colhead{$\bar E_{\nu_x}$} & & & \colhead{$L_{\bar\nu_e}$} &
 \colhead{$L_{\nu_x}$} & \\
 \colhead{Model} & \colhead{($M_\sun$)}& \colhead{(MeV)} &
 \colhead{(MeV)} & \colhead{$\beta_{\bar\nu_e}$} &
 \colhead{$\beta_{\nu_x}$} &
 \colhead{(ergs)} & \colhead{(ergs)} & \colhead{Reference}
}
\startdata
LL & 20 & 15.4 & 21.6 & 3.8 & 1.8 & $4.9\times 10^{52}$ &
 $5.0\times 10^{52}$ & 1 \\
TBP & 11 & 11.4 & 14.1 & 3.7 & 2.2 & \nodata & \nodata & 2
 \\
 & 15 & 11.4 & 14.1 & 3.7 & 2.2 & \nodata & \nodata & 2\\
 & 20 & 11.9 & 14.4 & 3.6 & 2.2 & \nodata & \nodata & 2\\
KRJ & \nodata & 15.4 & 15.7 & 4.2 & 2.5 & \nodata &
 \nodata & 3\\
\enddata
\tablerefs{(1) \citealt{Totani98}; (2) \citealt{Thompson03}; (3)
 \citealt{Keil03}.}
\end{deluxetable*}

Although the LL group succeeded in obtaining a robust explosion, their
result has recently been criticized because it lacked many relevant
neutrino processes that are now recognized as important.
Thus, we adopt the recent result of another hydrodynamic simulation, the
TBP one, which included all the relevant neutrino processes, such as
neutrino bremsstrahlung and neutrino-nucleon scattering with nucleon
recoil.
Their calculation obtained no explosion, and the neutrino spectrum ends
at 0.25 s after core bounce.
In the strict sense, we cannot use their result as our reference model
because the fully time-integrated neutrino spectrum is definitely
necessary in our estimate.
However, we adopt their result in order to confirm the effects of recent
sophisticated treatments of neutrino processes in the supernova core on
the SRN spectrum.
The TBP calculations include three progenitor mass models, i.e., 11, 15,
and 20$M_\sun$; all of these models are well fitted by equation
(\ref{eq:beta fit}), and the fitting parameters are summarized in Table
\ref{table:fitting parameters}.
The average energy for both $\bar\nu_e$ and $\nu_x$ is much smaller than
that by the LL calculation.
Although we do not show this in Table \ref{table:fitting parameters}, it
was also found that at least for the early phase of the core-collapse,
the complete equipartition of the gravitational binding energy for each
flavor was not realized.
However, it is quite unknown whether these trends hold during the entire
burst.
In this study, we adopt the average energy given in Table
\ref{table:fitting parameters} as our reference model, while we assume
perfect equipartition between flavors, i.e.,
$L_{\bar\nu_e}=L_{\nu_x}=5.0\times 10^{52}$ ergs.

In addition, we also use the model by KRJ.
Their calculation did not couple with the hydrodynamics, but it focused
on the spectral formation of neutrinos of each flavor using an MC
simulation.
Therefore, the static model was assumed as a background of neutrino
radiation, and we use their ``accretion phase model II,'' in which the
neutrino transfer was solved in the background of a 150 ms postbounce
model by way of a general relativistic simulation.
The fitting parameters for their MC simulation is also summarized in
Table \ref{table:fitting parameters}.
Unlike the previous two calculations, their result clearly shows that
the average energy of $\nu_x$ is very close to that of $\bar\nu_e$.
It also indicates that the equipartition among each flavor was not
realized, but rather $L_{\nu_e}\simeq L_{\bar\nu_e}\simeq 2L_{\nu_x}$.
However also in this case, since the totally time-integrated neutrino
flux is unknown from such temporary information, we assume perfect
equipartition, $L_{\bar\nu_e}=L_{\nu_x}=5.0\times 10^{52}$ ergs, as well
as that the average energies are the same as those in Table
\ref{table:fitting parameters}.

\subsection{Neutrino Spectrum after Neutrino Oscillation}
\label{sub:Neutrino Spectrum after Neutrino Oscillation}

The original $\bar\nu_e$ spectrum is different from what we observe as
$\bar\nu_e$ at Earth, owing to the effect of neutrino oscillation.
Since the specific flavor neutrinos are not mass eigenstates, they mix
with other flavor neutrinos during their propagation.
The behavior of flavor conversion inside the supernova envelope is well
understood, because the relevant mixing angles and mass square
differences are fairly well determined by recent solar, atmospheric, and
reactor neutrino experiments.
The remaining ambiguities concerning the neutrino oscillation parameters
are the value of $\theta_{13}$, which is only weakly constrained
($\sin^2\theta_{13}\lesssim 0.1$; \citealt{Apollonio99}), and the type
of mass hierarchy, i.e., normal ($m_1\ll m_3$) or inverted ($m_1\gg
m_3$).
We first discuss the case of normal mass hierarchy as our standard
model; in this case, the value of $\theta_{13}$ is irrelevant.
The case of inverted mass hierarchy is addressed in \S~\ref{sub:Inverted
Mass Hierarchy}.
In addition, other exotic mechanisms, such as resonant spin-flavor
conversion (see \citealt{Ando03g} and references therein) and neutrino
decay \citep{Ando03f}, which possibly change the SRN flux and spectrum,
might work in reality.
However, we do not consider such possibilities in this study.

The produced $\bar\nu_e$ at the supernova core are coincident with the
lightest mass eigenstate $\bar\nu_1$ owing to the large matter
potentials.
Since this state $\bar\nu_1$ is the lightest also in vacuum, there are
no resonance regions in which one mass eigenstate can change into
another state, and therefore $\bar\nu_e$ at production arrives at the
stellar surface as $\bar\nu_1$.
Thus, the $\bar\nu_e$ spectrum observed by the distant detector is
\begin{eqnarray}
 \frac{dN_{\bar\nu_e}}{dE_{\bar\nu_e}}
  &=&|U_{e1}|^2\frac{dN_{\bar\nu_1}}{dE_{\bar\nu_1}}
  +|U_{e2}|^2\frac{dN_{\bar\nu_2}}{dE_{\bar\nu_2}}
  +|U_{e3}|^2\frac{dN_{\bar\nu_3}}{dE_{\bar\nu_3}}\nonumber\\
  &=&|U_{e1}|^2\frac{dN_{\bar\nu_e}^0}{dE_{\bar\nu_e}}
   +\left(1-|U_{e1}|^2\right)\frac{dN_{\nu_x}^0}{dE_{\nu_x}},
   \label{eq:spectrum after oscillation}
\end{eqnarray}
where the quantities with superscript $0$ represent those at production,
$U_{\alpha i}$ is the mixing matrix element between the $\alpha$-flavor
state and $i$th mass eigenstate, and observationally $|U_{e1}|^2=0.7$.
In other words, 70\% of the original $\bar\nu_e$ survives; on the other
hand, the remaining 30\% comes from the other component $\nu_x$.
Therefore, both the original $\bar\nu_e$ and $\nu_x$ spectra are
necessary for the estimation of the SRN flux and spectrum; since the
original $\nu_x$ spectrum is generally harder than that of the original
$\bar\nu_e$, as shown in Table \ref{table:fitting parameters}, the
flavor mixing is expected to harden the detected SRN spectrum.

\section{FLUX AND EVENT RATE OF SUPERNOVA RELIC NEUTRINOS}
\label{sec:Flux and Event Rate of Supernova Relic Neutrinos}

\subsection{Flux of Supernova Relic Neutrinos}
\label{sub:Flux of Supernova Relic Neutrinos}

\begin{figure}[htbp]
\begin{center}
\includegraphics[width=8.5cm]{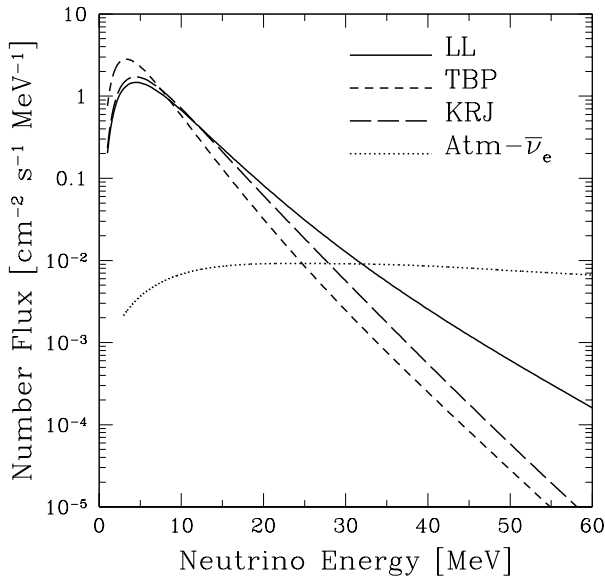}
\caption{SRN flux calculated with three reference models of original
 neutrino spectrum: LL \citep{Totani98}, TBP, and KRJ. The flux of
 atmospheric neutrinos \citep{Gaisser88,Barr89} is also shown for
 comparison. \label{fig:SRN_SNmodel}}
\end{center}
\end{figure}

The SRN flux can be calculated by equation (\ref{eq:SRN flux}) with our
reference models given in \S~\ref{sec:Formulation and Models}.
Figure \ref{fig:SRN_SNmodel} shows the SRN flux as a function of
neutrino energy for the three supernova models, LL, TBP, and KRJ.
The flux of atmospheric $\bar\nu_e$, which becomes background events for
SRN detection, is shown in the same figure
\citep*{Gaisser88,Barr89}.
The SRN flux peaks at $\lesssim 5$ MeV, and around this peak, the TBP
model gives the largest SRN flux because the average energy of the
original $\bar\nu_e$ is considerably smaller than in the other two
models but the total luminosity is assumed to be the same.
On the other hand, the model gives a smaller contribution at high-energy
regions, $E_\nu >10$ MeV.
In contrast, the high-energy tail of the SRN flux with the LL model
extends farther than with the other models, and it gives flux more than
1 mag larger at $E_\nu =60$ MeV.
This is because the high-energy tail was mainly contributed by the
harder component of the original neutrino spectrum; in the case of the
LL calculation, the average energy of the harder component $\nu_x$ is
significantly larger than that of the other two calculations, as shown
in Table \ref{table:fitting parameters}.
We show the values of the SRN flux integrated over the various energy
ranges in Table \ref{table:flux and event rate}.
\begin{deluxetable*}{crrrrrrr}
\tabletypesize{\scriptsize}
\tablecaption{Flux and Event Rate of Supernova Relic Neutrinos
 \label{table:flux and event rate}}
\tablehead{
 & & \multicolumn{3}{c}{Flux [cm${}^{-2}$ s${}^{-1}$]} &
 & \multicolumn{2}{c}{Event Rate [(22.5 kton yr)${}^{-1}$]} \\
 \cline{3-5} \cline{7-8}\\
 \colhead{Model} & Redshift Range & \colhead{Total} & \colhead{$E_\nu
 >11.3$ MeV} & \colhead{$E_\nu >19.3$ MeV} & & \colhead{$E_e >10$ MeV}
 & \colhead{$E_e >18$ MeV}
}
\startdata
LL & Total & 11.7 & 2.3 & 0.46 & & 2.3 & 1.0 \\
 & $0<z<1$\tablenotemark{a} & 4.1 (35.3) & 1.6 (70.9) & 0.39 (85.2) & &
 1.7 (77.5) & 0.9 (87.5) \\
 & $1<z<2$\tablenotemark{a} & 4.9 (42.0) & 0.6 (26.3) & 0.06 (14.0) & &
 0.5 (20.6) & 0.1 (11.9) \\
 & $2<z<3$\tablenotemark{a} & 1.8 (15.1) & 0.1 (2.5) & 0.0 (0.7) & & 0.0
 (1.7) & 0.0 (0.5) \\
 & $3<z<4$\tablenotemark{a} & 0.6 (5.3) & 0.0 (0.2) & 0.0 (0.0) & & 0.0
 (0.1) & 0.0 (0.0) \\
 & $4<z<5$\tablenotemark{a} & 0.2 (2.1) & 0.0 (0.0) & 0.0 (0.0) & & 0.0
 (0.0) & 0.0 (0.0) \\
TBP & Total & 16.1 & 1.3 & 0.14 & & 0.97 & 0.25 \\
KRJ & Total & 12.7 & 2.0 & 0.28 & & 1.7 & 0.53 \\
\cutinhead{Inverted Mass Hierarchy with Large $\theta_{13}$}
LL & Total & 9.4 & 3.1 & 0.94 & & 3.8 & 2.3 \\
TBP & Total & 13.8 & 1.9 & 0.30 & & 1.6 & 0.58 \\
KRJ & Total & 12.4 & 2.2 & 0.38 & & 2.0 & 0.76 \\
\enddata
\tablenotetext{a}{Contributions from each redshift range to the total
 ($0<z<5$) value are shown in parentheses as percentages.}
\tablecomments{Values in the upper part are evaluated for the case of
 normal mass hierarchy (or inverted mass hierarchy with sufficiently
 small $\theta_{13}$, i.e., $\sin^22\theta_{13}\lesssim 10^{-5}$), which
 we use as our standard model. On the other hand, values in the lower
 part are applicable only when the value of $\theta_{13}$ is large
 enough to induce completely adiabatic resonance, i.e.,
 $\sin^22\theta_{13}\gtrsim 10^{-3}$, in the case of inverted mass
 hierarchy.}
\end{deluxetable*}
(In the following, we refer only to the upper part of Table
\ref{table:flux and event rate}; the values in the lower part are
discussed in \S~\ref{sub:Inverted Mass Hierarchy}.)
The total flux is expected to be 11--16 cm${}^{-2}$ s${}^{-1}$ for our
reference models, although this value is quite sensitive to the shape of
the assumed SFR, especially at high-$z$.
The energy range in which we are more interested is high-energy regions
such as $E_\nu >19.3$ MeV and $E_\nu >11.3$ MeV, because as discussed
below, the background events are less critical and the reaction cross
section increases as $\propto E_\nu^2$.
In such a range, the SRN flux is found to be 1.3--2.3 cm${}^{-2}$
s${}^{-1}$ ($E_\nu >11.3$ MeV) and 0.14--0.46 cm${}^{-2}$ s${}^{-1}$
($E_\nu >19.3$ MeV).
Thus, the uncertainty about the supernova neutrino spectrum and its
luminosity gives at least a factor 2--4 ambiguity to the expected SRN
flux in the energy region of our interest.

\begin{figure}[htbp]
\begin{center}
\includegraphics[width=8.5cm]{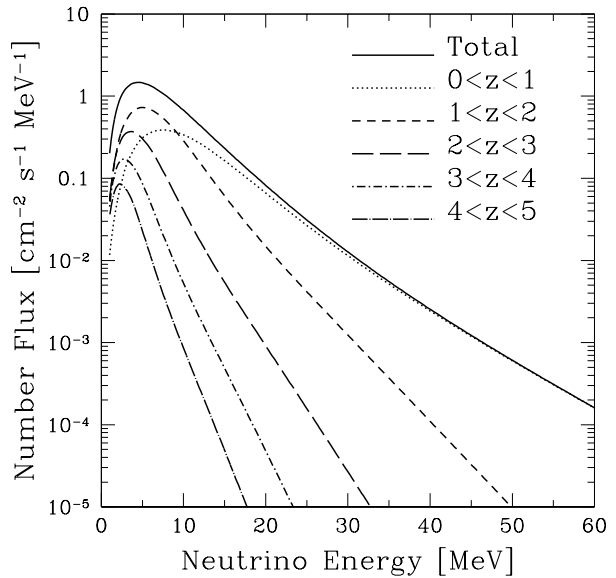}
\caption{SRN flux from various redshift ranges. LL is adopted as the
 supernova model. \label{fig:SRN_z}}
\end{center}
\end{figure}

Figure \ref{fig:SRN_z} shows the contribution by supernova neutrinos
emitted from various redshift ranges.
At high-energy region $E_\nu >10$ MeV, the dominant flux comes from the
local supernovae ($0<z<1$), while the low-energy side is mainly
contributed by the high-redshift events ($z>1$).
This is because the energy of neutrinos that were emitted from a
supernova at redshift $z$ is reduced by a factor of $(1+z)^{-1}$
reflecting the expansion of the universe, and therefore high-redshift
supernovae only contribute to low-energy flux.
We also show the energy-integrated flux from each redshift range in
Table \ref{table:flux and event rate} in the case of the LL supernova
model.
From the table, it is found that in the energy range of our interest,
more than 70\% of the flux comes from local supernova explosions at
$z<1$, while the high-redshift ($z>2$) supernova contribution is very
small.

\subsection{Event Rate at Water Cerenkov Detectors}
\label{sub:Event Rate at Water Cerenkov Detectors}

The water Cerenkov neutrino detectors have greatly succeeded in probing
the properties of neutrinos as elementary particles, such as neutrino
oscillation.
The SK detector is one of these detectors, and its large fiducial volume
(22.5 kton) might enable us to detect the diffuse background of SRNs.
Furthermore, much larger water Cerenkov detectors such as HK and UNO
are being planned.
SRN detection is most likely with the inverse $\beta$-decay reaction
with protons in water, $\bar\nu_ep\to e^+n$, and its cross section is
precisely understood \citep{Vogel99,Strumia03}.
In our calculation, we use the trigger threshold of SK-I (before the
accident).

\begin{figure}[htbp]
\begin{center}
\includegraphics[width=8.5cm]{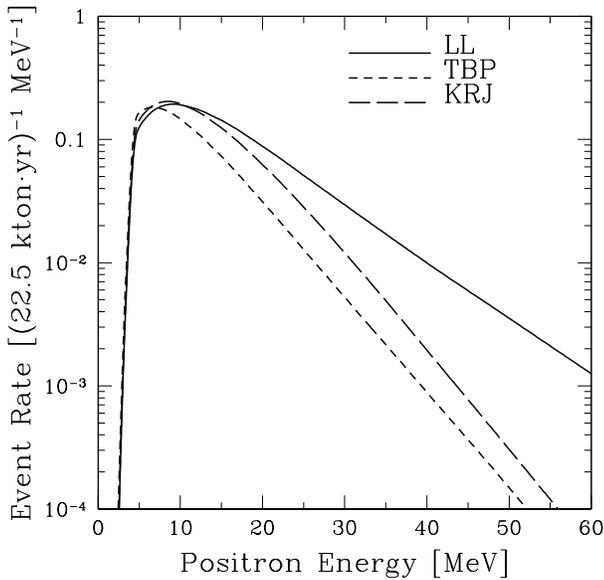}
\caption{Event rate at water Cerenkov detectors in units of (22.5 kton
 yr)${}^{-1}$ for three supernova models. \label{fig:evrt_SNmodel}}
\end{center}
\end{figure}

\begin{figure}[htbp]
\begin{center}
\includegraphics[width=8.5cm]{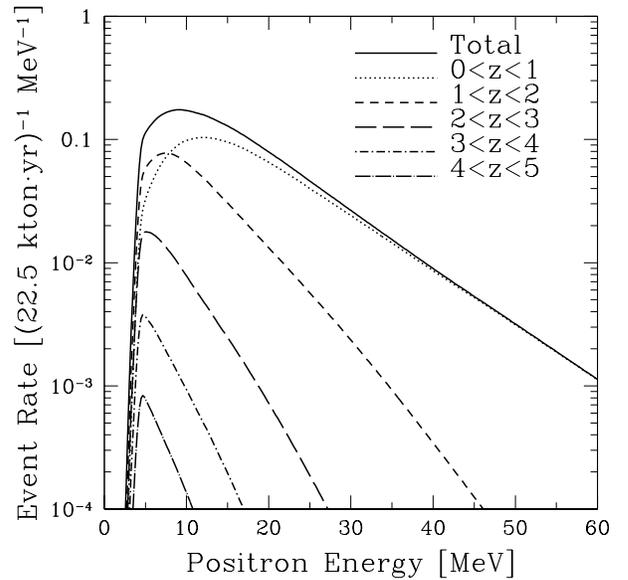}
\caption{Event rate at water Cerenkov detectors in units of (22.5 kton
 yr)${}^{-1}$ from various redshift ranges. LL is adopted as the
 supernova model. \label{fig:evrt_z}}
\end{center}
\end{figure}

The expected event rates at such detectors are shown in
Figures \ref{fig:evrt_SNmodel} and \ref{fig:evrt_z} in units of (22.5 kton
yr)${}^{-1}$ MeV${}^{-1}$; with SK, it takes a year to obtain the shown
SRN spectrum, while with HK and UNO, much less time [$1~{\rm yr}\times
(22.5~{\rm kton}/V_{\rm fid})$, where $V_{\rm fid}$ is the fiducial
volume of HK or UNO] is necessary because of their larger fiducial
volume.
Figure \ref{fig:evrt_SNmodel} compares the three models of the original
supernova neutrino spectrum, and Figure \ref{fig:evrt_z} shows the
contribution to the total event rate from each redshift range.
In Table \ref{table:flux and event rate} we summarize the event rate
integrated over various energy ranges for three supernova models.
The expected event rate is 0.97--2.3 (22.5 kton yr)${}^{-1}$ for $E_e
>10$ MeV and 0.25--1.0 (22.5 kton yr)${}^{-1}$ for $E_e>18$ MeV.
This clearly indicates that if the background events that hinder the
detection are negligible, the SK has already reached the required
sensitivity for detecting SRNs; with the future HK and UNO, a
statistically significant discussion would be possible.
This also shows that the current shortage of our knowledge concerning the
original supernova neutrino spectrum and luminosity gives at least a
factor of 2 ($E_\nu >10$ MeV) to 4 ($E_\nu >18$ MeV) uncertainty to the
event rate at the high-energy range (actual detection range).
We also summarize the contribution from each redshift range in the same
table, especially for the calculation with the LL model.
The bulk of the detected events will come from the local universe
$(z<1)$, but the considerable flux is potentially attributed to the
range $1<z<2$.

\subsection{Comparison with Other Studies and Current Observational
  Limits}
\label{sub:Comparison with Other Studies and Current Observational
Limit}

There are many past theoretical researches concerning the SRN flux
estimation based on a theoretically/observationally modeled cosmic SFR
\citep{Totani96a,Malaney97,Hartmann97,Ando03a}.
Here we briefly compare our results obtained in \S~\ref{sub:Flux of
Supernova Relic Neutrinos} and \S~\ref{sub:Event Rate at Water Cerenkov
Detectors} with these past analyses.
Our basic approach in the present paper is the same as that in
\citet{Ando03a}, in which the LL supernova model was adopted.
Thus the values for the LL model given in Table \ref{table:flux and
event rate} are almost the same as those in \citet{Ando03a}.
Two other studies \citep{Totani96a,Hartmann97} also used a similar
SFR-$z$ relation at low-redshift, and therefore their results are very
consistent with the present one (the LL model) at high-energy region
$E_\nu >10$ MeV.
Since the SFR model adopted by \citet{Malaney97} gave a rather lower
value at low-redshift, the resulting SRN flux at high-energy regions was
about a factor 2 smaller than our LL model or the other ones
\citep{Totani96a,Hartmann97,Ando03a}.
Thus, our calculation with the LL supernova model gives values quite
consistent with past studies within a factor of 2, but all of those
studies used the supernova model that is very similar to the LL model.
In fact, the present study is the first one to investigate the
dependence on the adopted supernova models, by using the various
original spectra of the different groups (LL, TBP, and KRJ).
As already mentioned in \S~\ref{sub:Flux of Supernova Relic Neutrinos},
it was found that the ambiguity concerning the original neutrino
spectrum varies the resulting value of the flux by at least a factor of
2.

In addition, there are several other studies on the SRN flux
\citep{Totani95,Kaplinghat00,Fukugita03}.
As for \citet{Totani95}, the authors used a constant supernova rate
model in order to investigate the dependence on cosmological parameters;
they gave a very large value ($\sim 3$ cm${}^{-2}$ s${}^{-1}$ at $E_\nu
>19.3$ MeV), which is already excluded observationally, because they
adopted a rather large supernova rate.
Concerning the other two studies, since neither of them gives a specific
value for the SRN flux, we cannot compare ours with theirs; they focused
on giving theoretical upper limit \citep{Kaplinghat00} or probing the
cosmic SFR with the current observational upper limit by SK
\citep{Fukugita03}.

Observationally, the SK collaboration gave a very stringent upper limit
to the SRN flux at $E_\nu >19.3$ MeV, i.e., 1.2 cm${}^{-2}$ s${}^{-1}$
(90\% CL; \citealt{Malek03}).
This number can be directly compared with our predictions summarized in
Table \ref{table:flux and event rate}.
Our predicted values are 0.46, 0.14, and 0.28 cm${}^{-2}$ s${}^{-1}$ for
the LL, TBP, and KRJ models, respectively.
Thus, the current SK upper limit is about a factor 2.5--8.5 larger than
our predictions with the reference model for the cosmic SFR, depending
on the adopted original neutrino spectrum.

\subsection{Background Events against Detection}
\label{sub:Background Events against Detection}

In \S~\ref{sub:Event Rate at Water Cerenkov Detectors} we calculated
the expected SRN spectrum at the water Cerenkov detectors on Earth, but
the actual detection is quite restricted because of the presence of
other background events.
There are atmospheric and solar neutrinos, antineutrinos from nuclear
reactors, spallation products induced by cosmic-ray muons, and decay
products of invisible muons (for a detailed discussion of these
backgrounds, see \citealt{Ando03a}).
For the pure-water Cerenkov detectors, there is no energy window in
which the flux of any backgrounds is much smaller than the SRN flux.

However, as proposed by \citet{Beacom03b}, if we use Gd-loaded
detectors, the range 10--30 MeV would be an energy window because we can
positively distinguish the $\bar\nu_e$ signal from other backgrounds
such as solar neutrinos ($\nu_e$), invisible muon events, and spallation
products; this is realized by capturing neutrons that are produced by
the $\bar\nu_ep$ interactions.
Above 30 MeV, the SRN flux becomes smaller than the flux of atmospheric
$\bar\nu_e$, as shown in Figure \ref{fig:SRN_SNmodel}; because they are
of the same flavor, it is in principle impossible to distinguish them
from the SRN $\bar\nu_e$.
On the other hand, below 10 MeV the reactor neutrinos ($\bar\nu_e$) are
the dominant background in the case of SK or HK; because the flux of
reactor neutrinos strongly depends on the detector site, it may be
possible to further reduce this lower energy cutoff (10 MeV) in the case
of UNO.

The neutron capture efficiency by Gd is estimated to be 90\% with the
proposed 0.2\% mixture by mass of GdCl$_3$ in water
\citep{Beacom03b}, and subsequently 8 MeV gamma cascade occurs from the
excited Gd.
The single-electron energy equivalent to this cascade was found to be
3--8 MeV by careful simulation \citep{Hargrove95}, and with the
trigger threshold adopted in SK-I, only about 50\% of such cascades can
be detected actually.
However, it is expected that SK-III, which will begin operation in
mid-2006, will trigger at 100\% efficiency above 3 MeV, with good
trigger efficiency down to 2.5 MeV \citep{Beacom03b}.
In that case most of the gamma cascades from Gd will be detected with
their preceding signal of positrons.
From this point on, we assume 100\% efficiency; even if we abandon this
assumption, it does not affect our physical conclusion, since the
relevant quantity representing the detector performance is (fiducial
volume) $\times$ (time) $\times$ (efficiency), which we call effective
volume.

\section{MONTE CARLO SIMULATION FOR FUTURE DETECTOR PERFORMANCE}
\label{sec:Monte Carlo Simulation for Future Detector Performance}

In this section we predict the expected signal at future detectors, such
as Gd-SK, Gd-HK, and Gd-UNO, using the MC method with our reference
models.
These pseudodata are then analyzed using several free parameters
concerning the SFR.
Although we focus on how far the SFR can be probed by SRN observation,
the uncertainty from the supernova neutrino spectrum would give a fair
amount of error.
However, this problem can be solved if a supernova explosion occurs in
our Galaxy; the expected event number is about 5000--10,000 at SK, when
supernova neutrino burst occurs at 10 kpc, and it will enable a
statistically significant discussion concerning the neutrino spectrum
from supernova explosions.
Even if there are no Galactic supernovae in the near future, remarkable
development of the supernova simulation can be expected with the growth
of computational resources and numerical technique.
With such developments, the supernova neutrino spectrum and luminosity
may be uncovered, and the ambiguity is expected to be reduced
significantly.
Thus, in this paper we assume that the supernova neutrino spectrum is
well understood and that our reference models are fairly good
representatives of nature; we analyze the SFR alone with several free
parameters.

The basic procedure of our method is as follows.
(1) We simulate the expected signal (spectrum) at a Gd-loaded detector
in the range 10--30 MeV, assuming that there are no background events.
In that process, we use our reference models for the generation of the
SRN signal (eq. [\ref{eq:SFR}] for the SFR and the LL model as neutrino
spectrum).
(2) Then we analyze the SRN spectrum using the maximum likelihood method
with two free parameters of the SFR and obtain a set of the best-fit
values for those parameters; they are concerned with the supernova
rate as
\begin{equation}
R_{\rm SN}(z)=\left\{
	       \begin{array}{ccc}
		R_{\rm SN}^0 (1+z)^\alpha & {\rm for} & z<1,\\
		2^\alpha R_{\rm SN}^0 & {\rm for} & z>1,
\end{array}\right.
\end{equation}
where $R_{\rm SN}^0$ represents the local supernova rate and $\alpha$
determines the slope of supernova rate evolution.
Although it is recognized that the SFR-$z$ relation increases from $z=0$
to $z=1$ from various observations using light, the actual numbers
for the absolute value and the slope of the SFR-$z$ relation are still a
matter of controversy and independent confirmation, such as ours, is
needed.
We assume that the comoving SFR is constant at $z>1$; even if we changed
this assumption, the result would be the same because the bulk of the
detected event comes from local supernovae.
(3) We perform 10$^3$ such MC simulations and obtain 10$^3$ independent
sets of best-fit parameters.
Then we discuss the standard deviation of the distributions of such
best-fit parameter sets and the implications for the cosmic SFR.

\subsection{Performance of the Gd-loaded Super-Kamiokande Detector}
\label{Performance of the Gd-loaded Super-Kamiokande Detector}

\begin{figure}[htbp]
\begin{center}
\includegraphics[width=8.5cm]{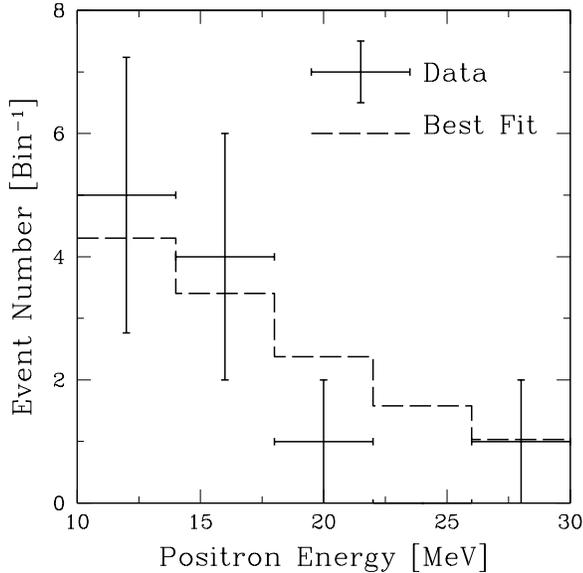}
\caption{Expected SRN spectrum at a detector of effective volume
 22.5 kton\,$\times$\,5 yr. The data points represent the result of MC
 simulation, and the error bars include statistical errors alone. These
 data generated by MC simulation were analyzed assuming $R_{\rm
 SN}^0=1.2\times 10^{-4}$ yr${}^{-1}$ Mpc${}^{-3}$, and using $\alpha$
 as a free parameter. The best-fit value for $\alpha$ is 2.97 and it
 resulted in dashed histogram in this figure. \label{fig:MC_spectrum}}
\end{center}
\end{figure}

\begin{figure}[htbp]
\begin{center}
\includegraphics[width=8.5cm]{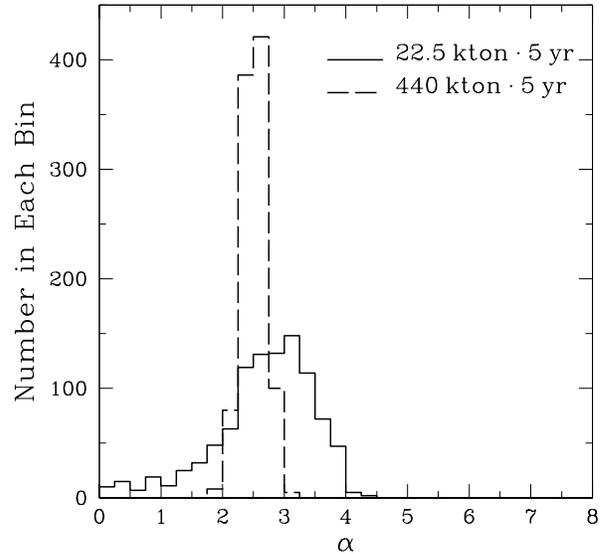}
\caption{Distribution of 10$^3$ best-fit values for $\alpha$, which are
 obtained from the analyses of each MC generation. The effective volume
 is 22.5 kton\,$\times$\,5 yr for solid histogram, and 440
 kton\,$\times$\,5 yr for dashed histogram. The value of the local
 supernova rate is fixed to be $R_{\rm SN}^0=1.2\times 10^{-4}$
 yr${}^{-1}$ Mpc${}^{-3}$. \label{fig:MC_hist_index}}
\end{center}
\end{figure}

In this subsection we discuss the performance of Gd-SK for 5 years, or
an effective volume of $22.5 ~{\rm kton}\times 5 ~{\rm yr}$.
Because the expected event number is only $\sim 10$, the parameters
$R_{\rm SN}^0$ and $\alpha$ cannot both be well determined at once.
Therefore, we fix one of those parameters with some value inferred from
other observations.
First, the value of $R_{\rm SN}^0$ was fixed to be $1.2\times 10^{-4}$
yr${}^{-1}$ Mpc${}^{-1}$, which was inferred from the local supernova
survey \citep{Madau98b}, and we obtained the distribution of the
best-fit values of parameter $\alpha$.
Figure \ref{fig:MC_spectrum} shows the expected SRN spectrum; points
with error bars represent the result of one MC simulation, and the
dashed histogram is the spectrum with the best-fit parameter ($\alpha
=2.97$).
Thus, from one realization of the MC simulation, we obtain one best-fit
parameter.
The result of 10$^3$ MC simulations are shown in
Figure \ref{fig:MC_hist_index} as a histogram of the distribution of
best-fit parameters $\alpha$ ({\it solid histogram}).
The average value of these 10$^3$ values for $\alpha$ is found to be
2.67, and the standard deviation is 0.80, i.e., $\alpha =2.67\pm 0.80$.
A no-evolution (constant supernova rate) model would be excluded at the
$3.3\sigma$ level from the SRN observation alone with an effective
volume of $22.5 ~{\rm kton}\times 5 ~{\rm yr}$.

\begin{figure}[htbp]
\begin{center}
\includegraphics[width=8.5cm]{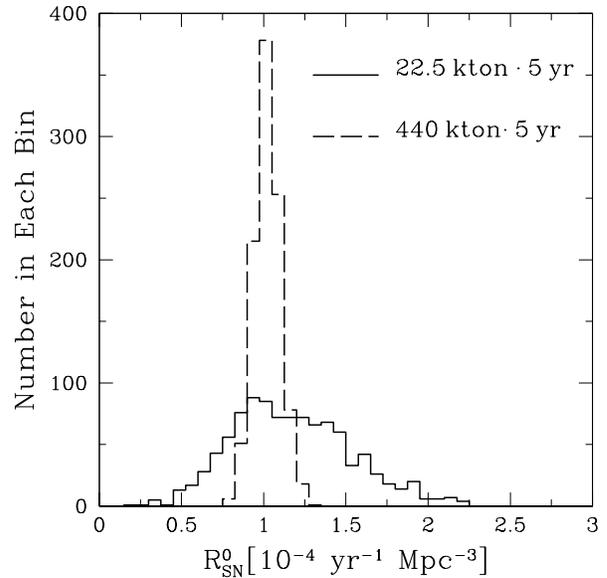}
\caption{Distribution of 10$^3$ best-fit values for local supernova rate
 $R_{\rm SN}^0$, which are obtained from the analyses of each MC
 generation. The effective volume is 22.5 kton\,$\times$\,5 yr for solid
 histogram, and 440 kton\,$\times$\,5 yr for dashed histogram. The value
 of $\alpha$ is fixed to be 2.9. \label{fig:MC_hist_norm}}
\end{center}
\end{figure}

Then in turn, we fixed the value of $\alpha$ to be 2.9 in order to
obtain the distribution of best-fit values for the local supernova rate
$R_{\rm SN}^0$ from the SRN observation.
The result of 10$^3$ MC generations and analyses in this case is shown in
Figure \ref{fig:MC_hist_norm}.
The average value for $R_{\rm SN}^0$ is $1.2\times 10^{-4}$ yr${}^{-1}$
Mpc${}^{-3}$, and the standard deviation is $0.4\times 10^{-4}$
yr${}^{-1}$
Mpc${}^{-3}$, i.e., $R_{\rm SN}^0=(1.2\pm 0.4)\times 10^{-4}$
yr${}^{-1}$ Mpc${}^{-3}$.

\begin{figure}[htbp]
\begin{center}
\includegraphics[width=8.5cm]{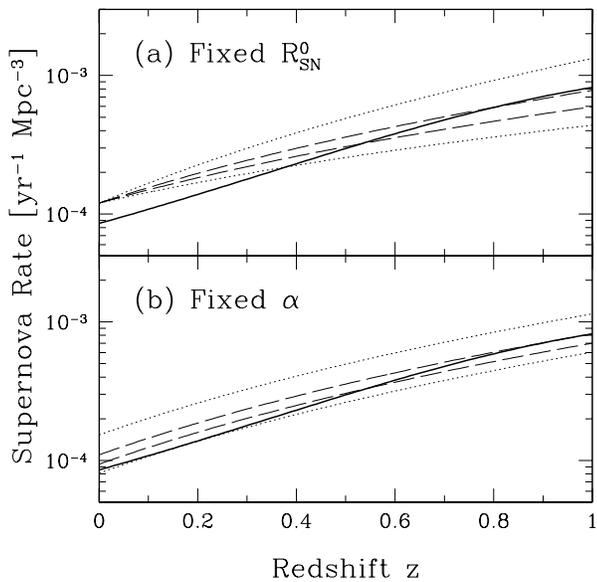}
\caption{Supernova rate as a function of redshift. In both panels, solid
 curves represent our reference model. ({\it a}) The allowed region at
 the $1\sigma$ level, concerning the fitting parameter $\alpha$ with
 fixed $R_{\rm SN}^0$, is shown as the area between the two dotted
 curves for an effective volume of $22.5 ~{\rm kton}\times 5 ~{\rm yr}$
 and as the area between the two dashed curves for an effective volume
 of $440 ~{\rm kton} \times 5 ~{\rm yr}$. ({\it b}) Same as ({\it
 a}) but for fitting parameter $R_{\rm SN}^0$ with fixed
 $\alpha$. \label{fig:SNrate}}
\end{center}
\end{figure}

The results obtained with the above calculations are summarized in Table
\ref{table:expected parameter values}.
\begin{deluxetable*}{cccrcrc}
\tabletypesize{\scriptsize}
\tablecaption{Expected Sensitivity of Future Detectors to Supernova Rate
 Model \label{table:expected parameter values}}
\tablehead{
 & \colhead{Effective Volume} & \colhead{Fixed} &
 & \colhead{$\delta\alpha/\langle\alpha\rangle$} &
 \colhead{$R_{\rm SN}^0$} & \colhead{$\delta R_{\rm SN}^0/\langle R_{\rm
 SN}^0\rangle$}\\
 \colhead{Detector} & \colhead{(22.5 kton yr)} & \colhead{Parameter} &
 \colhead{$\alpha$} & (\%) & \colhead{(10${}^{-4}$ yr${}^{-1}$
 Mpc${}^{-3}$)} & (\%)
}
\startdata
 SK & 5 & $R_{\rm SN}^0$ & $2.7\pm 0.8$ & 30.0 & 1.2 (fixed) & \nodata\\
 & 5 & $\alpha$ & 2.9 (fixed) & \nodata & $1.2\pm 0.4$ & 28.3\\
 HK or UNO & 97.8 & $R_{\rm SN}^0$ & $2.5\pm 0.2$ & 7.8 & 1.2 (fixed)
 & \nodata\\
 & 97.8 & $\alpha$ & 2.9 (fixed) & \nodata & $1.0\pm 0.1$ & 7.7\\
 & 97.8 & \nodata & $3.5\pm 1.3$ & 36.7 & $0.88\pm 0.48$ & 54.8\\
\enddata
\end{deluxetable*}
In Figure \ref{fig:SNrate} we compare the supernova rate model in which
the parameter is inferred from the MC simulations with the ``true''
reference model; the cases of fixed $R_{\rm SN}^0$ and $\alpha$ are
shown in Figures \ref{fig:SNrate}{\it a} and \ref{fig:SNrate}{\it b},
respectively.
The allowed region at the $1\sigma$ level is located between the two
dotted curves, while the solid curve represents our reference model.
Thus, with the Gd-SK detector we can roughly reproduce the supernova
rate profile at $z<1$ for 5 years operation, although it is still
statistically insufficient.

\subsection{Future Gd-loaded Megaton-Class Detectors}
\label{sub:Future Gd-loaded Mega-ton Class Detectors}

We consider future megaton-class detectors such as Gd-HK or Gd-UNO.
With these detectors, the effective volume that we consider, $440 ~{\rm
kton}\times 5 ~{\rm yr}$, is expected to be realized in several years
from the start of their operation.
First we did the same analysis adopted in the previous subsection, i.e.,
we fixed one of relevant parameters, $\alpha$ or $R_{\rm SN}^0$, and
investigated the dependence on the remaining parameter.
The values that we used for fixed parameters were the same as those
given in the previous subsection.
The result of these cases are also shown in
Figures \ref{fig:MC_hist_index} and \ref{fig:MC_hist_norm} as dashed
histograms, which give $\alpha =2.51\pm 0.20$ and $R_{\rm SN}^0=(1.0\pm
0.1)\times 10^{-4}$ yr${}^{-1}$ Mpc${}^{-3}$, respectively, and these
values are also summarized in Table \ref{table:expected parameter
values}.
The statistical errors are considerably reduced compared with the case
of 22.5 kton\,$\times$\,5 yr, because of the $\sim 20$ times larger
effective volume.
Thus, future magaton detectors will possibly pin down, within 10\%
statistical error, either the index of supernova rate evolution $\alpha$
or the local supernova rate $R_{\rm SN}^0$ if the other is known in
advance.
The dashed curves in Figure \ref{fig:SNrate} set the allowed region of
the supernova rate at the $1\sigma$ level by the considered detectors,
well reproducing the assumed model.

\begin{figure}[htbp]
\begin{center}
\includegraphics[width=8.5cm]{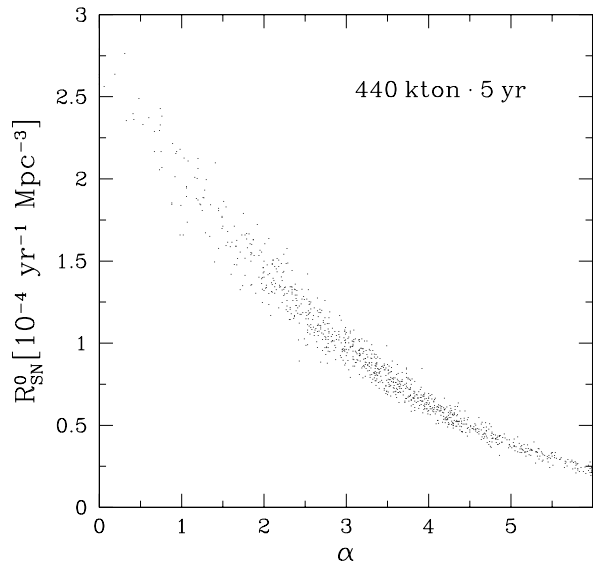}
\caption{Distribution of 10$^3$ best-fit parameter sets ($\alpha,
 R_{\rm SN}^0$). Each dot represents the result of one MC
 generation and an accompanying analysis. The effective volume is 440
 kton\,$\times$\,5 yr. \label{fig:MC_index_norm}}
\end{center}
\end{figure}

In principle, we can determine both parameters by SRN observation,
because $R_{\rm SN}^0$ is concerned with the absolute value of the flux
alone but $\alpha$ is concerned with both the absolute value and the
spectral shape; i.e., these two parameters are not degenerate with each
other.
Thus, we repeated the same procedure but without fixing the values of
$\alpha$ or $R_{\rm SN}^0$.
The distribution of 10$^3$ best-fit parameter sets of ($\alpha,R_{\rm
SN}^0$) is shown in Figure \ref{fig:MC_index_norm} for a detector with
an effective volume of 440 kton\,$\times$\,5 yr; the mean values and the
standard deviations are $\alpha =3.5\pm 1.3$ and $R_{\rm SN}^0=(8.8\pm
4.8)\times 10^{-5}$ yr${}^{-1}$ Mpc${}^{-3}$.
Even though the effective volume is as large as 440 kton\,$\times$\,5
yr, it is still insufficient for determining both parameters at once.
For another trial, we also carried out the same MC simulations, but
using a hypothetical (and unrealistic) effective volume as large as 440
kton\,$\times$\,10$^4$ yr.
In that case the free parameters are found to be quite well constrained
at $\alpha =3.68\pm 0.03$ and $R_{\rm SN}^0=(7.11\pm 0.09)\times
10^{-5}$ yr${}^{-1}$ Mpc${}^{-3}$.

\section{DISCUSSION}
\label{sec:Discussion}

\subsection{Supernova Rate at High-Redshift Region}
\label{sub:Supernova Rate at High Redshift}

At the detection energy range 10--30 MeV that we have considered, the
main contribution to the SRN event rate comes from low-redshift region
$0<z<1$, as shown in Figure \ref{fig:evrt_z} and Table
\ref{table:flux and event rate}.
However, if we can reduce the lower energy threshold $E_{\rm th}$, we
expect that the contribution of supernova neutrinos from high-redshift
$z>1$ becomes enhanced.
The value of $E_{\rm th}$ is restricted to 10 MeV because at energy
regions lower than this, there is a large background of reactor
neutrinos; its removal is impossible with the current detection
methods.
Since the SK and HK detectors are and will be located at Kamioka in
Japan, they are seriously affected by background neutrinos from many
nuclear reactors.
If some large-volume detectors were built at a location free from such
background, the lower threshold energy could be reduced, enabling us
to probe the high-redshift supernova rate.
In this subsection, thus, we discuss the detector performance as a
function of the value of $E_{\rm th}$.

\begin{figure}[htbp]
\begin{center}
\includegraphics[width=8.5cm]{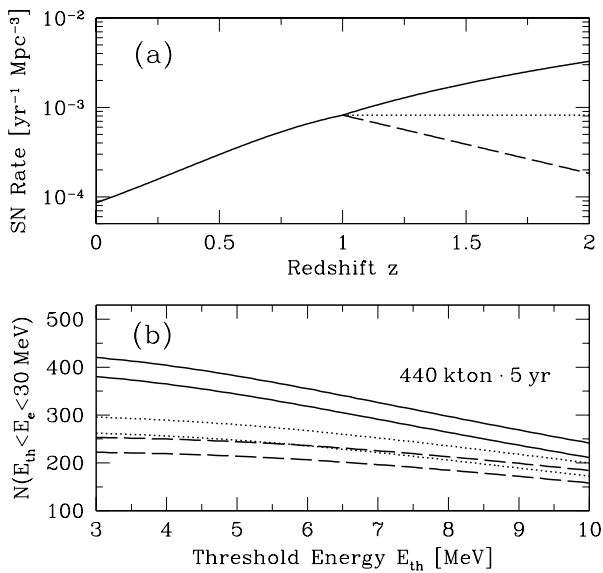}
\caption{({\it a}) Three toy models for comoving density of supernova
 rate as a function of redshift. ({\it b}) Expected event number $N$ at
 $E_{\rm th}< E_e<30$ MeV as a function of $E_{\rm th}$, for the
 fiducial volume of 440 kton\,$\times$\,5 yr. The line types correspend
 to those in ({\it a}). The upper and lower curves of each type
 represent $N+\sqrt N$ and $N-\sqrt N$, respectively; i.e., the area
 between the two curves is the allowed region at the $1\sigma$
 level. \label{fig:N_Eth_relation}}
\end{center}
\end{figure}

In Figure \ref{fig:N_Eth_relation}{\it a} we show three toy models of
comoving density of supernova rate as a function of redshift.
These models exactly coincide with each other at $z<1$ (and also with
the previous reference model represented by eq. [\ref{eq:SFR}]) but
seriously differ at $z>1$.
We calculate the expected event number for $E_{\rm th}<E_e<30$ MeV at a
detector with an effective volume of 440 kton\,$\times$\,5 yr, using
these toy models, the LL spectrum, and trigger threshold expected at
SK-III.
The result is shown in Figure \ref{fig:N_Eth_relation}{\it b}.
As expected, the discrepancy among the three models becomes larger as we
reduce the threshold energy.
In particular, the model that produces larger numbers of supernovae at
$z>1$ ({\it solid curve}) is satisfactorily distinguishable in the case
of sufficiently low $E_{\rm th}$.
This is because the larger supernova rate at high-redshift region $z>1$
increases the fraction of its contribution to the SRN flux.
On the other hand, the model with a lower supernova rate relatively
increases the contribution from low-redshift region $z<1$, and therefore
the difference between constant ({\it dotted curve}) and decreasing
model ({\it dashed curve}) is less prominent.

\subsection{Probing Supernova Neutrino Properties}
\label{sub:Probing Supernova Neutrino Properties}

Until this point, we have assumed that the properties of supernova
neutrinos, such as the average energy difference between flavors and
luminosities of neutrinos of different flavors, will be quite well
understood when future SRN detection comes within reach.
However, this assumption itself is quite unclear because Galactic
supernova explosions, which would give us rich information on the
supernova neutrino spectrum and luminosity, may not occur by the time we
are ready for the SRN detection.
Furthermore, there is no assurance that the numerical experiments will
succeed in obtaining the supernova explosion itself and predicting the
supernova neutrino properties precisely by then.
Thus, SRN observation might be the only probe of the supernova neutrino
properties.

In this subsection, we discuss how far we can derive the supernova
neutrino properties from SRN observation.
We have already shown that even using data of 440 kton\,$\times$\,5 yr,
at most only two free parameters can be satisfactorily constrained.
Therefore, we now have to adopt another assumption such that the
evolution of the supernova rate is quite well understood by future
observations with the various planned satellites and telescopes.
The procedure is basically the same as that of the previous section;
i.e., we run 10$^3$ MC simulations and analyze these psudodata to obtain
the best-fit values for two free parameters, $\bar E_{\bar\nu_e}$ and
$\bar E_{\nu_x}/\bar E_{\bar\nu_e}$.
The values of $\beta_\nu$ and $L_{\nu}$ defined in equation
(\ref{eq:beta fit}) are assumed to be $\beta_{\bar\nu_e}=4.0$,
$\beta_{\nu_x}=2.2$, and $L_{\bar\nu_e}=L_{\nu_x}=5.0\times 10^{52}$
ergs.
As a result of such calculations, we obtain the distribution of the
two parameters, which is characterized by $E_{\bar\nu_e}=(15.9\pm 1.3)$
MeV and $\bar E_{\nu_x}/\bar E_{\bar\nu_e}=1.5\pm 0.4$; although this
well reproduces the LL model, the errors are still very large.
Considering that many uncertainties concerning the SFR estimate possibly
remain even in future updated observations, the errors to these
quantities would be much larger than the purely statistical ones given
above.

\subsection{Inverted Mass Hierarchy}
\label{sub:Inverted Mass Hierarchy}

Throughout the above discussion, we have assumed normal hierarchy of
neutrino masses ($m_1\ll m_3$).
However, the case of inverted mass hierarchy has not been experimentally
excluded yet, and we explore this possibility in this subsection.
In this case, flavor conversions inside the supernova envelope change
dramatically, compared with the normal mass hierarchy already discussed
in \S~\ref{sub:Neutrino Spectrum after Neutrino Oscillation}.
Since $\bar\nu_3$ is the lightest, $\bar\nu_e$ are created as
$\bar\nu_3$, owing to large matter potential.
In that case, it is well known that at a so-called resonance point,
there occurs a level crossing between $\bar\nu_1$ and $\bar\nu_3$ (for a
more detailed discussion, see, e.g., \citealt{Dighe00}).
At this resonance point, complete $\bar\nu_1\leftrightarrow\bar\nu_3$
conversion occurs when the so-called adiabaticity parameter is
sufficiently small compared to unity (it is said that resonance is
``nonadiabatic''), while conversion never occurs when it is large
(adiabatic resonance).
The adiabaticity parameter $\gamma$ is quite sensitive to the value of
$\theta_{13}$, i.e., $\gamma\propto\sin^22\theta_{13}$; when
$\sin^22\theta_{13}\gtrsim 10^{-3}$ ($\sin^22\theta_{13}\lesssim
10^{-5}$), the resonance is known to be completely adiabatic
(nonadiabatic) \citep{Dighe00}.
When the resonance is completely nonadiabatic (because of small
$\theta_{13}$), the situation is the same as in the case of normal mass
hierarchy already discussed in \S~\ref{sub:Neutrino Spectrum after
Neutrino Oscillation} (because $\bar\nu_e$ at production become
$\bar\nu_1$ at the stellar surface), and the $\bar\nu_e$ spectrum after
oscillation is represented by equation (\ref{eq:spectrum after
oscillation}).
On the other hand, adiabatic resonance (due to large $\theta_{13}$)
forces $\bar\nu_e$ at production to become $\bar\nu_3$ when they escape
from the stellar surface, and therefore the observed $\bar\nu_e$
spectrum is given by
\begin{equation}
 \frac{dN_{\bar\nu_e}}{dE_{\bar\nu_e}}
  =|U_{e3}|^2\frac{dN_{\bar\nu_e}^0}{dE_{\bar\nu_e}}
   +\left(1-|U_{e3}|^2\right)\frac{dN_{\nu_x}^0}{dE_{\nu_x}}
   \simeq\frac{dN_{\nu_x}^0}{dE_{\nu_x}}.
   \label{eq:spectrum after oscillation for inverted hierarchy}
\end{equation}
The second equality follows from the fact that the value of $|U_{e3}|^2$
is constrained to be much smaller than unity from reactor experiments
\citep{Apollonio99}.
Thus, equation (\ref{eq:spectrum after oscillation for inverted
hierarchy}) indicates that complete conversion takes place between
$\bar\nu_e$ and $\nu_x$.
When the value of $\theta_{13}$ is large enough to induce adiabatic
resonance ($\sin^22\theta_{13}\gtrsim 10^{-3}$), the obtained SRN flux
and spectrum should be very different from ones obtained in
\S\S~\ref{sub:Flux of Supernova Relic Neutrinos} and \ref{sub:Event
Rate at Water Cerenkov Detectors}.
The SRN flux and event rate in this case were calculated with equations
(\ref{eq:SRN flux}) and (\ref{eq:spectrum after oscillation for inverted
hierarchy}), and the results are summarized in the lower part of Table
\ref{table:flux and event rate}.
The values (with the LL model) shown in this table are consistent with
the previous calculation by \citet{Ando03c}, in which numerically
calculated conversion probabilities were adopted with some specific
oscillation parameter sets (which include a model with inverted mass
hierarchy and $\sin^22\theta_{13}=0.04$), as well as realistic stellar
density profiles.

The total flux becomes 9.4--14 cm${}^{-2}$ s${}^{-1}$, somewhat smaller
than the values given in the upper part of the same table, because the
total flux is dominated by the low-energy region.
The fluxes at $E_\nu >19.3$ MeV are enhanced to be 0.30--0.94
cm${}^{-2}$ s${}^{-1}$, but this is still below the current 90\% CL
upper limit of $1.2$ cm${}^{-2}$ s${}^{-1}$ obtained by the SK
observation.
The event rate at the future detectable energy range, $E_\nu >10$ MeV,
is expected to become 1.6--3.8 yr${}^{-1}$, which is considerably larger
than the values in the case of normal mass hierarchy, 0.97--2.3
yr${}^{-1}$.
The increase (decrease) of the flux and event rate integrated over the
high (total) energy region is due to the very high efficiency of the
flavor conversion, $\nu_x\to\bar\nu_e$, inside the supernova envelope;
because the original $\nu_x$ are expected to be produced with larger
average energy as shown in Table \ref{table:fitting parameters}, the
efficient conversion makes the SRN spectrum harder, which enhances the
flux and event rate at the high-energy region.
Thus, if the inverted mass hierarchy, as well as the large value for
$\theta_{13}$, were realized in nature, SRN detection would be rather
easier, compared with the other cases.
Although we do not repeat the MC simulations that were introduced in
\S~\ref{sec:Monte Carlo Simulation for Future Detector Performance}, the
results can be easily inferred; the statistical errors in this case
would be $\sim (3.8/2.3)^{1/2}=1.3$ times smaller than the values given
in Table \ref{table:expected parameter values}, because they are
inversely proportional to the square root of the event number.

\section{CONCLUSIONS}
\label{sec:Conclusions}

In the present paper, we have investigated the flux and event rate of
SRNs and discussed their implications for the cosmic SFR.
Since SRNs are diffuse neutrino background emitted from past
core-collapse supernova explosions, they contain fruitful information
not only on the supernova neutrino spectrum itself but also on the
supernova rate in the past and present universe, which is quite
difficult to estimate because, e.g., the problem of dust extinction is
nontrivial.
As reference models, we adopted the supernova rate model based on recent
SFR observations (eq. [\ref{eq:SFR}]) and the supernova neutrino
spectrum numerically calculated by three groups (LL, TBP, and KRJ).
As a result of our calculations, the flux integrated over the entire
energy region was found to be 12--16 cm${}^{-2}$ s${}^{-1}$, depending
on the adopted supernova neutrino spectrum (Table \ref{table:flux and
event rate}).
Although there is no energy window for the SRN detection at present
owing to various background events, in the near future, it is expected
that the energy region of 10--30 MeV will be utilized for SRN
detection.
This is due to the technique of neutron capture by dissolved Gd.
In the detection energy range $E_e>10$ MeV, the SRN event rate was found
to be 0.97--2.3 yr${}^{-1}$ at a detector with a fiducial volume of 22.5
kton (Table \ref{table:flux and event rate}).

We also simulated the expected signal with one set of the reference
models by using the Monte Carlo method and then analyzed these
pseudodata with several free parameters, obtaining one set of best-fit
values for them.
MC simulations repeated $10^3$ times gave 10$^3$ independent best-fit
parameter sets, and we gave a statistical discussion using their
distribution.
First of all, we used parameterization such that $R_{\rm SN}(z)=R_{\rm
SN}^0 (1+z)^\alpha$, where $R_{\rm SN}^0$ and $\alpha$ are free
parameters, assuming that the supernova neutrino spectrum and luminosity
are well understood by way of a future Galactic supernova neutrino burst
or future development of the numerical supernova simulations.
The obtained distribution for these two parameters was found to be
represented by $\alpha =2.7\pm 0.8, \delta\alpha/\langle\alpha\rangle
=30\%$ and $R_{\rm SN}^0 =(1.2\pm 0.4) \times 10^{-4} ~{\rm yr}^{-1}
~{\rm Mpc}^{-3}, \delta R_{\rm SN}^0/\langle R_{\rm SN}^0\rangle =28\%$
for a detector with an effective volume of 22.5 kton\,$\times$\,5 yr,
and $\alpha =2.5\pm 0.2, \delta\alpha/\langle\alpha\rangle =7.8\%$ and
$R_{\rm SN}^0 =(1.0\pm 0.1) \times 10^{-4} ~{\rm yr}^{-1} ~{\rm
Mpc}^{-3}, \delta R_{\rm SN}^0/\langle R_{\rm SN}^0\rangle =7.7\%$ for a
detector with an effective volume of 440 kton\,$\times$\,5 yr, where one
of the parameters is fixed (Figs. \ref{fig:MC_hist_index} and
\ref{fig:MC_hist_norm}; Table \ref{table:expected parameter values}).
The parameterized supernova rate models with the obtained parameter
values are compared with the assumed reference model in Figure
\ref{fig:SNrate}, and we found that the fitting model well reproduced
the reference model.
On the other hand, if we fix neither value for these two parameters,
the expected errors become rather large at $\delta\alpha /\langle
\alpha\rangle =37\%$ and $\delta R_{\rm SN}^0/\langle R_{\rm SN}^0
\rangle =55\%$, even with an effective volume of 440 kton\,$\times$\,5
yr.

In addition, we explored several other possibilities in
\S~\ref{sec:Discussion}.
First, we discussed the dependence of the event number on the adopted
lower cutoff energy.
Although below 10 MeV there is a background of reactor neutrinos, their
flux strongly depends on the detector sites, and the lower energy
threshold $E_{\rm th}$ could possibly be reduced.
We investigated the expected event number for $E_{\rm th}<E_e<30$ MeV as
a function of $E_{\rm th}$ in Figure \ref{fig:N_Eth_relation} for various
toy models of supernova rate and found that the model that produces
larger number of supernovae at $z>1$ is satisfactorily distinguishable
in the case of sufficiently small $E_{\rm th}$.
Second, the SRN spectrum as a potential probe of the supernova
neutrino spectrum itself was investigated, because such an approach
might be very important if there are no Galactic supernova explosions
in the near future or no successful numerical supernova simulations.
We discussed using the same MC procedure, but assuming that the
supernova rate is quite well understood.
Although the obtained distribution reproduces properties of the LL
spectrum, the errors were still found to be large, and considering the
uncertainties concerning the SFR, these errors are only lower limits;
the actual errors would be much larger.
Finally, the case of an inverted mass hierarchy was investigated.
We showed that only in the case in which $\sin^22\theta_{13}\gtrsim
10^{-5}$ the values of the SRN flux should be modified.
The results in the case of completely adiabatic resonance, which is
realized when $\sin^22\theta_{13}\gtrsim 10^{-3}$, are shown in the
lower part of Table \ref{table:flux and event rate}.
In this case, it was found that the expected event rate would be
enhanced to 1.6--3.8 yr${}^{-1}$, although these values are still
below the current upper bound; SRN detection would be more probable
in this case.

\acknowledgments

This work was supported by a Grant-in-Aid for JSPS Fellows.

\bibliography{refs}

\end{document}